\documentclass[fleqn,twoside]{article}
\usepackage{espcrc2}

\usepackage{graphicx}

\title{Constraints on the Neutrino Mass from Cosmology and their impact on 
world neutrino data.}
\author{A.\ Melchiorri\address{Dipartimento di Fisica and Sezione INFN, Universit\`a degli Studi di Roma ``La Sapienza'', P.le Aldo Moro 5, 00185, Rome,
Italy}, 
G.L.\ Fogli, E.\ Lisi, A.\ Marrone, A. \ Palazzo\address{Dipartimento di Fisica 
and Sezione INFN di Bari, Via Amendola 173, 70126, Bari, Italy},
P.\ Serra\address{Dipartimento di Fisica, Universit\`a degli Studi di 
Roma ``La Sapienza'', P.le Aldo Moro 5, 00185, Rome, Italy} 
and J.\ Silk\address{Astrophysics, Denys Wilkinson Building, Keble Road, OX13RH,Oxford, United Kingdom}}

\begin{document}
%\date{{\today}}
\begin{abstract}
We derive upper limits on the sum of neutrino masses 
from an updated combination of data from Cosmic Microwave
Background experiments and Galaxy Redshifts Surveys. 
The results are discussed in the context of three-flavor neutrino mixing
and compared with neutrino oscillation data, with upper limits on 
the effective neutrino mass in Tritium beta decay from the Mainz and Troitsk 
experiments and with the claimed lower bound on the effective Majorana 
neutrino mass in neutrinoless double beta decay from the Heidelberg-Moscow 
experiment.
\end{abstract}

\maketitle

%%%%%%%%%%%%%%%%%%%%%%%%%%%%%%%%%%%%%%%%%%%%%%%%%%%%%%%%%%%%%%%%%%%%%%
%%%% Section I %%%%%%%%%%%%%%%%%%%%%%%%%%%%%%%%%%%%%%%%%%%%%%%%%%%%%%%
%%%%%%%%%%%%%%%%%%%%%%%%%%%%%%%%%%%%%%%%%%%%%%%%%%%%%%%%%%%%%%%%%%%%%%

\section{Introduction}

Cosmological observations have started to provide valuable upper limits 
on absolute neutrino masses (see, e.g., the reviews \cite{Barg,Dolg}), 
competitive with those from laboratory experiments. 
In particular, the combined analysis of
high-precision data from Cosmic Microwave Background (CMB)
anisotropies and Large Scale Structures (LSS) has already reached a
sensitivity of $O(\mathrm{eV})$ (see, e.g., \cite{Be03,Tg04,Laha})
for the sum of the neutrino masses $\Sigma$,
%...................................................................
\begin{equation}\label{Sigma}
\Sigma = m_1+m_2+m_3\ .
\end{equation}
%...................................................................
We recall that the total neutrino energy density in our Universe,
$\Omega_{\nu}h^2$ (where $h$ is the Hubble constant normalized to
$H_0=100$ km~s$^{-1}$~Mpc$^{-1}$) is related to $\Sigma$ by the
well-known relation $\Omega_{\nu}h^2=\Sigma / (93.2 \mathrm{\ eV})$
\cite{PDG4}, and plays an essential role in theories of structure
formation. It can thus leave key signatures in LSS data 
(see, eg.,\cite{Hu98}) and, to a lesser extent, in CMB data 
(see, e.g.,\cite{Ma95}). Very recently, it has also been shown that accurate
Lyman-$\alpha$ (Ly$\alpha$) forest data \cite{Mc04}, taken at face
value, can improve the current CMB+LSS constraints on $\Sigma$ by a
factor of $\sim 3$, with important consequences on absolute neutrino
mass scenarios\cite{Se04}. 

On the other hand, atmospheric, solar, reactor 
and accelerator neutrino experiments have convincingly established 
that neutrinos are massive and mixed. 
World neutrino data are consistent with a three-flavor mixing
framework (see \cite{fogli04} and references therein), parameterized in
terms of three neutrino masses $(m_1,m_2,m_3)$ and of three mixing
angles $(\theta_{12},\theta_{23},\theta_{13})$, plus a
possible CP violating phase $\delta$.

Neutrino oscillation experiments are sensitive to two independent
squared mass difference, $\delta m^2$ and $\Delta m^2$ (with $\delta
m^2\ll \Delta m^2$), hereafter defined as \cite{Delt}
%..................................................................
\begin{equation}
\label{DeltaDef} (m^2_1,m^2_2,m^2_3)= \mu^2 + \left( -\frac{\delta
m^2}{2}, +\frac{\delta m^2}{2},\pm\Delta m^2 \right),
\end{equation}
%..................................................................
where $\mu$ fixes the absolute neutrino mass scale, while the cases
$+\Delta m^2$ and $-\Delta m^2$ identify the so-called normal and
inverted neutrino mass hierarchies, respectively. Neutrino
oscillation data indicate that $\delta m^2\simeq 8\times 10^{-5}$
eV$^2$ and $\Delta m^2\simeq 2.4\times 10^{-3}$ eV$^2$. 
They also indicate that $\sin^2\theta_{12}\simeq
0.3$, $\sin^2\theta_{23}\simeq 0.5$,
and $\sin^2\theta_{13}\leq \mathrm{few}\%$. However,
they are currently unable to determine the mass hierarchy
($\pm\Delta m^2$) and the phase $\delta$, and are
insensitive to the absolute mass parameter $\mu$ in
Eq.~(\ref{DeltaDef}).

The absolute neutrino mass scale can also be probed by non-oscillatory
neutrino experiments. The most sensitive laboratory experiments to
date have been focussed on tritium beta decay 
and on neutrinoless double beta decay. Beta
decay experiments probe the so-called effective electron neutrino
mass $m_\beta$ \cite{Mbet},
%....................................................................
\begin{equation}
\label{mb} m_\beta =
\left[c^2_{13}c^2_{12}m^2_1+c^2_{13}s^2_{12}m^2_2+s^2_{13}m^2_3
\right]^\frac{1}{2}\ ,
\end{equation}
%....................................................................
where $c^2_{ij}=\cos^2\theta_{ij}$ and $s^2_{ij}=\sin^2\theta_{ij}$.
Current experiments (Mainz \cite{Main} and Troitsk \cite{Troi})
provide upper limits in the range $m_\beta\leq \mathrm{few}$~eV
\cite{PDG4,Eite}.

Neutrinoless double beta decay ($0\nu2\beta$) experiments are
instead sensitive to the so-called effective Majorana mass
$m_{\beta\beta}$ (if neutrinos are Majorana fermions),
%.................................................................
\begin{equation}\label{mbb}
m_{\beta\beta} = \left|
c^2_{13}c^2_{12}m_1+c^2_{13}s^2_{12}m_2e^{i\phi_2}+s^2_{13}m_3
e^{i\phi_3}\right|\ ,
\end{equation}
%...................................................................
where $\phi_2$ and $\phi_3$ parameterize relative (and unknown)
Majorana neutrino phases \cite{ScVa}. All $0\nu2\beta$ experiments
place only upper bounds on $m_{\beta\beta}$ (the most sensitive
being in the eV range, with the exception of the
Heidelberg-Moscow experiment \cite{Kl01}, which claims a positive
(but highly debated) $0\nu2\beta$ signal $m_{\beta\beta}>0.17$ eV at
$95 \%$ c.l. and corresponding to $m_{\beta\beta}$ in the sub-eV range 
at best fit \cite{Kl03,Kl04}.

In these proceedings, we will briefly illustrate the impact of
the cosmological constraints on the sum of neutrino masses on the 
three-flavor mixing theoretical and observational scenario. 

%In \cite{fogli04} we performed a global phenomenological analysis of the
%constraints applicable to the observables
%$(m_\beta,m_{\beta\beta},\Sigma)$, by using up-to-date experimental
%data and state-of-the-art calculations for all the relevant
%laboratory and astrophysical quantities.

\section{Upper bounds on $\Sigma$ from cosmological data}

The neutrino contribution to the overall energy density of the
universe can play a relevant role in large scale structure formation
and leave key signatures in several cosmological data sets. More
specifically, neutrinos suppress the growth of fluctuations on
scales below the horizon when they become non relativistic. A
massive neutrinos of a fraction of eV would therefore produce a
significant suppression in the clustering on small cosmological
scales (namely, for comoving wavenumber $k\sim 0.05 \ h\
\mathrm{Mpc}^{-1}$).

To constrain $\Sigma$ from cosmological data, we perform a
likelihood analysis comparing the recent observations with a set of
models with cosmological parameters sampled as follows: cold dark
matter (cdm) density $\Omega_\mathrm{cdm}h^2 \in [0.05,0.20]$ in
steps of $0.01$; baryon density $\Omega_{b}h^2 \in [0.015, 0.030]$
(motivated by Big Bang Nucleosynthesis) in steps of $0.001$; a
cosmological constant $\Omega_{\Lambda} \in [0.50, 0.96]$ in steps
of  $0.02$; and neutrino density $\Omega_{\nu}h^2 \in [0.001,
0.020]$ in steps of $0.002$. We restrict our analysis to {\it flat}
$\Lambda$-CDM models, $\Omega_\mathrm{tot}=1$, and we add a
conservative external prior on the age of the universe, $t_0 > 10$
Gyrs. 
The value of the Hubble constant in our database is not an
independent parameter, since it is determined through the flatness
condition. We adopt the conservative top-hat bound $0.50 < h < 0.90$
and we also consider the $1\sigma$ constraint on the Hubble
parameter, $h=0.71\pm0.07$, obtained from Hubble Space Telescope
(HST) measurements~\cite{freedman}. We allow for a reionization of
the intergalactic medium by varying the CMB photon optical depth
$\tau_c$ in the range $\tau_c \in [0.05,0.30]$ in steps of $0.02$.

We restrict the analysis to adiabatic inflationary models with a
negligible contribution of gravity waves. We let vary the spectral index $n$
of scalar primordial fluctuations in the range $n\in [0.85, 1.3]$
and its running $dn/d\ln k \in [-0.40,0.2]$ assuming pivot scales at
$k_0=0.05 \mathrm{\ Mpc}^{-1}$ and $k_0=0.002 \mathrm{\ Mpc}^{-1}$.
We rescale the fluctuation amplitude by a prefactor $C_{110}$, in
units of the value $C_{110}^\mathrm{WMAP}$ measured by the Wilkinsin
Microwave Anisotropy Probe (WMAP) satellite. Finally, concerning the
neutrino parameters, we fix the number of neutrino species to
$N_{\nu}=3$, all with the same mass (the effect of mass differences
compatible with neutrino oscillation being negligible in the current
cosmological data \cite{pastor}). An higher number of neutrino
species can weakly affect both CMB and LSS data (see, e.g.,
\cite{bowen}) but is highly constrained by standard big bang
nucleosynthesis and is not considered in this work, where we focus
on $3\nu$ mixing.

The cosmological data we considered comes from observation of CMB
anisotropies and polarization, galaxy redshift surveys and
luminosity distances of type Ia supernovae. For the CMB data we use
the recent temperature and cross polarization results from the WMAP
satellite \cite{Be03} using the method explained in \cite{map5} and
the publicly available code.
Given a theoretical temperature anisotropy and polarization angular
power spectrum in our database, we can therefore associate a
$\chi^2_\mathrm{WMAP}$ to the corresponding theoretical model.

We further include the latest results from other CMB datasets.
The CMB data analysis methods have been already described in
\cite{fogli04} and will not be reported here.

In addition to the CMB data we also consider the real-space power
spectrum of galaxies from either the 2 degrees Fields (2dF) Galaxy
Redshifts Survey or the Sloan Digital Sky Survey (SDSS), using the
data and window functions of the analysis of \cite{thx} and
\cite{Tg04}. We restrict the analysis to a range of scales over
which the fluctuations are assumed to be in the linear regime ($k <
0.2 h^{-1}\mathrm{\ Mpc}$). When combining with the CMB data, we
marginalize over a bias $b$ for each data set considered as an
additional free parameter.

We also include information from the Ly$\alpha$ Forest in the SDSS,
using the results of the analysis of \cite{Se04} and \cite{Mc04},
which probe the amplitude of linear fluctuations at very small
scales. For this data set, small-scale power spectra are computed at
high redshifts and compared with the values presented in
\cite{Mc04}. As in \cite{Se04}, we do not consider running.

We finally incorporate constraints obtained from the SN-Ia
luminosity measurements of \cite{riess} using the so-called GOLD
data set. Luminosity distances at SN-Ia redshifts are computed for
each model in our database and compared with the observed apparent
bolometric SN-Ia luminosities.
In Fig.~1 we plot the likelihood distribution for $\Sigma$ from our
joint analysis of CMB~+~SN-Ia~+~HST~+~LSS data, transformed into an
equivalent $\Delta\chi^2_\Sigma$ function, which allows to derive
bounds on $\Sigma$ at any fixed confidence level. We take LSS data
either from the SDSS or the 2dF survey (dashed and solid curves,
respectively).%
%----------------------------------
\footnote{For the sake of brevity, the subdominant block of data
(SN-Ia~+~HST) is not explicitly indicated in figure labels.}
%---------------------------------

As we can see, these curves do not show evidence for a neutrino mass
(the best fit being at $\Sigma\simeq 0$) and provide the $2\sigma$
bound $\Sigma \leq 1.4$ eV. Such bound is in good agreement with
previous results in similar analyses
\cite{Be03,hannestad,Tg04,barger,crotty}. 

Also plotted in Fig.~1 is the  $\Delta\chi^2_\Sigma$ function from a
joint analysis of CMB~+~SN-Ia~+~HST~+~2dF~+~Ly$\alpha$. No running
is assumed in this analysis, and we find a $2\sigma$ bound $\Sigma <
0.47$ eV, in very good agreement (despite the more approximate
method we used) with the analysis already presented in \cite{Se04}.

As shown in Fig.~1 and already discussed in \cite{Se04}, the
inclusion of the Ly$\alpha$ data from the SDSS set greatly improves
the constraints on $\Sigma$.

%---------------------------------------------------------------------------
\begin{figure}
\includegraphics[width=6.0cm]{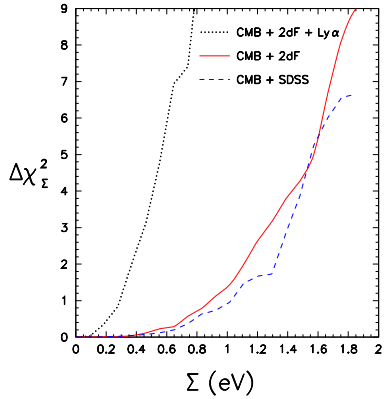}
\caption{\label{fig2} Upper bounds on the sum of
neutrino masses $\Sigma$ from our $3\nu$ analysis of cosmological
data, given in terms of the $\Delta\chi^2_\Sigma$ function. The
solid and dashed curves refer to the combination of  CMB and LSS
data (CMB+2dF and CMB+SDSS, respectively). The two CMB+LSS fits
provide comparable results and, for definiteness, the CMB+2df one is
adopted. In addition, we consider also the case where the recent
Ly$\alpha$ data from the SDSS are included, providing significantly
stronger constraints on $\Sigma$ (dotted curve). See \cite{fogli04} for
details.}
\end{figure}
%---------------------------------------------------------------------------

%---------------------------------------------------------------------------
\begin{figure}
\includegraphics[width=6.0cm]{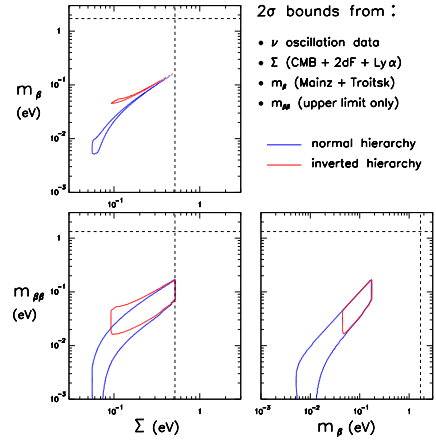}
\caption{\label{fig4} Global $3\nu$ analysis in the
$(m_{\beta},m_{\beta\beta},\Sigma)$ parameter space, using
oscillation data plus laboratory data and cosmological  data. 
This figure implements also upper limits (shown
as dashed lines at $2\sigma$ level) on $m_\beta$ from Mainz+Troitsk
data, on $m_{\beta\beta}$ from $0\nu2\beta$ data, and on $\Sigma$
from CMB+2dF+Ly$\alpha$ data. In combination with oscillation parameter bounds,
the cosmological upper limit on $\Sigma$ dominates over the
laboratory upper limits on $m_\beta$ and $m_{\beta\beta}$.
See \cite{fogli04} for details.}
\end{figure}
%---------------------------------------------------------------------------

\section{Adding bounds from laboratory and Astrophysics}

Here we consider confidence regions obtained 
from analysis of neutrino oscillation data, of 
$m_\beta$  and  $m_{\beta\beta}$ data and cosmological CMB+LSS 
data (see \cite{fogli04} for more details). 

Figure~2 shows such regions projected in the three coordinate planes.
Separate laboratory and cosmological upper bounds at the $2\sigma$
level are shown as dashed lines, while the regions allowed by the
combination of laboratory, cosmological, and oscillation data are
shown as thick solid curves for normal hierarchy and as thin solid
curves for inverted hierarchy. It can be seen that the upper bounds
on the $(m_\beta,m_{\beta\beta},\Sigma)$ observables are dominated
by the cosmological upper bound on $\Sigma$. This bound, via the
$(m_\beta,\Sigma)$ and $(m_{\beta\beta},\Sigma)$ correlations
induced by oscillation data, provides upper limits also on
$m_{\beta\beta}$ and $m_\beta$, which happen to be stronger than the
current laboratory limits by a factor $\sim 4$. 

Since significant 
improvements on laboratory limits for $m_{\beta\beta}$ and $m_\beta$
will require new experiments and several years of data taking
\cite{Eite}, cosmological determinations of $\Sigma$, although
indirect, will continue to provide, in the next future, the most
sensitive upper limits (and hopefully a signal) for absolute
neutrino mass observables.

In Fig.~2, the tension (at $2\sigma$) 
between the limits from cosmology and the lower limit on  
$m_{\beta\beta} >0.17$ eV claimed by the Heidelberg-Moscow
experiment is a clear symptom of possible problems, either in some 
data sets or in their theoretical interpretation, which definitely 
prevent any global combination of data. 
It would be premature to conclude that, e.g., the $0\nu2\beta$ 
claim is``ruled out'' by cosmological data but it is anyway exciting 
that global neutrino data analyses have already 
reached a point where fundamental questions may start to arise.

{\it Acknowledgements}
A.M. would like to thank the Organizers of the 
{\em NOW-2004\/} workshop. The work of A.M.
is supported by the Italian Ministero dell'Istruzione, Universit\`a 
e Ricerca (MIUR) through the ``GEMINI'' research project
and Istituto Nazionale di Fisica Nucleare (INFN) through the
``Astroparticle Physics'' research project. 
The work of G.L.F., E.L., A.M.$2$, and A.P.\ is supported by the
Italian Ministero dell'Istruzione, Universit\`a e Ricerca (MIUR) and
Istituto Nazionale di Fisica Nucleare (INFN) through the
``Astroparticle Physics'' research project.

%\end{document}

\end{document}